\documentclass[sigconf]{acmart}

\AtBeginDocument{%
  }

\usepackage{tcolorbox}

\usepackage{bbding}
\usepackage{subfigure}
\usepackage[boxed,ruled,lined]{algorithm2e}
\usepackage{multirow}
\usepackage{makecell}

\usepackage{arydshln} 
\usepackage{enumerate}
\usepackage{enumitem}
\usepackage{makecell}
\usepackage{bbm}

\newcommand{\ourname}{LLaDA-Rec\xspace}

\setcopyright{acmlicensed}
\copyrightyear{2018}
\acmYear{2018}
\acmDOI{XXXXXXX.XXXXXXX}
\acmConference[Conference acronym 'XX]{Make sure to enter the correct
  conference title from your rights confirmation email}{June 03--05,
  2018}{Woodstock, NY}
\acmISBN{978-1-4503-XXXX-X/2018/06}

\begin{document}

\title{\ourname: Discrete Diffusion for Parallel Semantic ID Generation in Generative Recommendation}

\author{Teng Shi}
\author{Chenglei Shen}
\affiliation{
\institution{\mbox{Gaoling School of Artificial Intelligence}\\Renmin University of China}
  \city{Beijing}
  \country{China}
}
\email{{shiteng,chengleishen9}@ruc.edu.cn}

\author{Weijie Yu}
\affiliation{
\institution{School of Information Technology and Management\\University of International Business and Economics}
  \city{Beijing}
  \country{China}
}
\email{yu@uibe.edu.cn}

\author{Shen Nie}
\author{Chongxuan Li}
\author{Xiao Zhang}
\affiliation{
\institution{\mbox{Gaoling School of Artificial Intelligence}\\Renmin University of China}
  \city{Beijing}
  \country{China}
}
\email{{nieshen,chongxuanli,zhangx89}@ruc.edu.cn}

\author{Ming He}
\affiliation{
\institution{AI Lab at Lenovo Research}
  \city{Beijing}
  \country{China}
}
\email{heming01@foxmail.com}

\author{Yan Han}
\affiliation{
\institution{AI Lab at Lenovo Research}
  \city{Beijing}
  \country{China}
}
\email{hanyan7217@163.com}

\author{Jun Xu}
\affiliation{
\institution{\mbox{Gaoling School of Artificial Intelligence}\\Renmin University of China}
  \city{Beijing}
  \country{China}
}
\email{junxu@ruc.edu.cn}

\renewcommand{\shortauthors}{Teng Shi et al.}

\begin{abstract}
Generative recommendation represents each item as a semantic ID, i.e., a sequence of discrete tokens, and generates the next item through autoregressive decoding.
While effective, existing autoregressive models face two intrinsic limitations: 
(1)~unidirectional constraints, where causal attention restricts each token to attend only to its predecessors, hindering global semantic modeling; and 
(2)~error accumulation, where the fixed left-to-right generation order causes prediction errors in early tokens to propagate to the predictions of subsequent token.
To address these issues, we propose \textbf{LLaDA-Rec}, a \emph{discrete diffusion} framework that reformulates recommendation as parallel semantic ID generation. 
By combining bidirectional attention with the adaptive generation order, the approach models inter-item and intra-item dependencies more effectively and alleviates error accumulation. 
Specifically, our approach comprises three key designs: 
(1)~a parallel tokenization scheme that produces semantic IDs for bidirectional modeling, addressing the mismatch between residual quantization and bidirectional architectures; 
(2)~two masking mechanisms at the user-history and next-item levels to capture both inter-item sequential dependencies and intra-item semantic relationships; and 
(3)~an adapted beam search strategy for adaptive-order discrete diffusion decoding, resolving the incompatibility of standard beam search with diffusion-based generation.
Experiments on three real-world datasets show that LLaDA-Rec consistently outperforms both ID-based and state-of-the-art generative recommenders, establishing discrete diffusion as a new paradigm for generative recommendation.
\end{abstract}



\maketitle

\section{Introduction}
In recent years, generative language models~\cite{achiam2023gpt,zhao2023survey,yang2025qwen3,liu2024deepseek} have exhibited remarkable capabilities, motivating a growing number of studies to explore the application of the generative paradigm to recommendation~\cite{TIGER,LETTER,LC_Rec,RPG,zhou2025onerec,deng2025onerec}. In contrast to traditional discriminative recommendation models~\cite{SASREC,BERT4REC,FMLPREC}, generative recommendation reformulates the recommendation task as a sequence-to-sequence generation problem, where the next target item is generated in an autoregressive manner.

Generative recommendation typically comprises two key components: item tokenization and autoregressive generation.
In item tokenization, each item is assigned 
a Semantic ID 
(SID),
represented as a sequence of discrete tokens. Common techniques include hierarchical clustering~\cite{si2024generative} (e.g., K-means) and vector quantization (e.g., RQ-VAE~\cite{TIGER,LC_Rec}). The user history encoded as discrete tokens is subsequently fed into the generative model, which autoregressively produces the tokens of the next item in a step-by-step manner.

\begin{figure}[t]
\centering
\includegraphics[width=0.95\columnwidth]{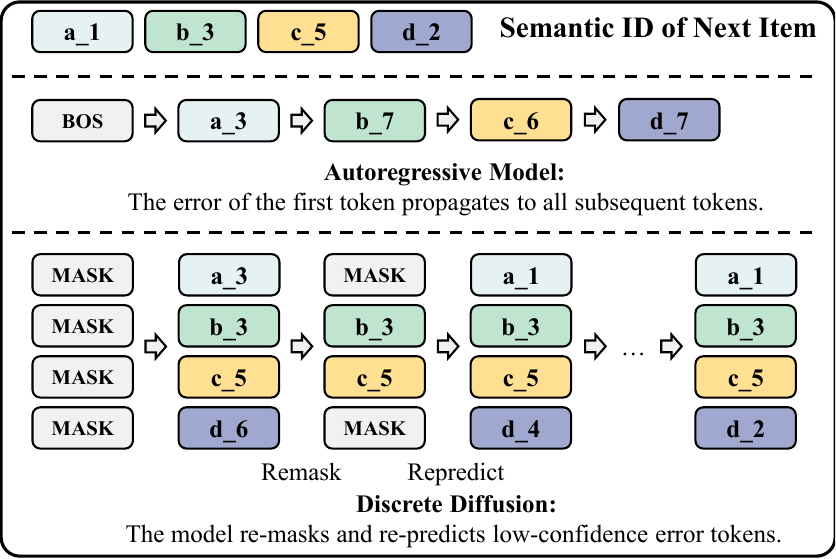}
\caption{
An illustration of the advantages of discrete diffusion over autoregressive generation. 
In autoregressive models, an error in the first token will propagate to subsequent tokens. 
In contrast, a discrete diffusion model predicts all masked positions in parallel at each step, 
and re-masks and re-predicts low-confidence error tokens, ultimately producing more accurate results.
}
\label{fig:intro_example}
\vspace{-0.3cm}
\end{figure}

Despite their effectiveness, existing methods still exhibit several limitations:
(1)~\textbf{Unidirectional Constraint}. 
Most approaches rely on autoregressive architectures that predict the tokens of each item in a left-to-right manner. 
This left-to-right dependency restricts the model’s ability to capture global relationships among all tokens that jointly define an item~\cite{gu2022vector}. As a result, the generated items often exhibit limited semantic coherence and expressiveness.
(2)~\textbf{Error Accumulation}. During inference, models generate each token conditioned on previously sampled ones, in contrast to the training phase where teacher forcing~\cite{raffel2020exploring} is employed and the ground-truth token is provided at each step. 
As a result, errors made on earlier tokens during inference cannot be corrected and are likely to propagate to subsequent tokens, thereby amplifying their adverse effects~\cite{lin2025order}.
Figure~\ref{fig:intro_example} provides a further illustration.

To address these limitations, we explore a new formulation of generative recommendation based on discrete diffusion~\cite{li2025survey,nie2025large,zhu2025llada,yang2025mmada}.
During training, the model employ a forward token masking noise process followed by a reverse denoising reconstruction process, with a bidirectional Transformer parameterizing the prediction of masked tokens.
This design offers two key advantages:
(1)~\textbf{Bidirectional Understanding}.
The bidirectional architecture enhances contextual comprehension and enables the modeling of global relationships among all tokens.
(2)~\textbf{Adaptive Generation Order}.
During inference, all tokens are predicted in parallel, while low‑confidence tokens are iteratively re‑masked and re‑predicted. This yields a dynamic rather than fixed generation order, prioritizing tokens with higher model confidence (i.e., easier tokens) and mitigating error accumulation.

Overall, this discrete diffusion framework offers a compelling solution to the limitations identified in prior generative recommendation approaches. Although diffusion models inherently operate under a fixed‑length generation constraint~\cite{nie2025large}, this property naturally aligns with recommendation tasks, where each item is represented by a fixed number of semantic IDs—making diffusion particularly well‑suited for generative recommendation.

However, applying discrete diffusion to generative recommendation still presents several challenges:
(1)~\textbf{Mismatch between Residual Quantization (RQ) and Discrete Diffusion.}
Since discrete diffusion relies on a bidirectional Transformer where all tokens are equally important, it requires parallel tokenization schemes beyond hierarchical ones like RQ-VAE.
(2)~\textbf{Beam Search is Not Directly Applicable to Discrete Diffusion.}
Unlike standard discrete diffusion models that rely on probabilistic sampling, recommendation tasks require generating the top-$k$ outputs. 
Beam search, which is widely used in autoregressive top-$k$ generation~\cite{TIGER,LC_Rec}, is primarily designed for fixed left-to-right decoding and thus cannot be directly applied to discrete diffusion with an adaptive generation order, thereby requiring specialized decoding strategies tailored to the discrete diffusion setting.
(3)~\textbf{Differences between Language Modeling and Recommendation.}
Original discrete diffusion models are primarily developed for language modeling. When applied to recommendation tasks, they need to be adapted to capture intra‑item semantics and inter‑item sequential relations within user histories.

To address the above challenges, we propose \textbf{\ourname}, a bidirectional generative recommendation method based on discrete diffusion.
Our approach consists of three main modules:
(1)~\textbf{Parallel Tokenization}. We design 
Multi-Head VQ-VAE
that encode the item embedding and split it into several sub‑vectors, each queried against a separate codebook, ultimately producing parallel semantic IDs.
(2)~\textbf{Discrete Diffusion Training}. We employ two masking mechanisms: a User‑History level masking to model inter‑item sequential relationships, and a Next‑Item level mask, with the history fixed, to capture intra‑item semantics, enabling the model to better understand the semantic relationships among different tokens within an item. 
(3)~\textbf{Discrete Diffusion Inference}. We adapt beam search for diffusion model. At each step, we first select the position with the highest model confidence for generation, and then perform beam expansion at that position. Through multiple iterations, this process produces the final top-$k$ recommended~items.

This paper makes the following key contributions:

\begin{itemize}[leftmargin=*]
    \item 
    We analyze the issues of unidirectional constraints and error accumulation in existing autoregressive generative recommendation models, which limit their capability.
    \item We propose \ourname, a generative recommendation model based on discrete diffusion, which introduces parallel semantic IDs and develops discrete diffusion training and inference methods tailored to recommendation tasks.
    \item Experiments on three datasets demonstrate the effectiveness of \ourname, which consistently outperforms both traditional item-ID-based methods and semantic-ID-based generative recommendation models, thereby establishing discrete diffusion as a promising new paradigm for generative recommendation.
\end{itemize}

\section{Related Work}

\subsection{Generative Recommendation}
Inspired by the success of large language models (LLMs)~\cite{zhao2023survey}, generative recommendation has garnered growing attention~\cite{zhai2024actions,ETEGRec,hua2023index,TIGER,LC_Rec,LETTER,RPG,deng2025onerec,zhou2025onerec,shi2025gensar}. In this paradigm, each item is tokenized into a sequence of discrete tokens, known as semantic IDs (SIDs). Consequently, every item in a user’s interaction history is represented by its SIDs and fed into a generative model to produce the SIDs of the target item.
In general, generative recommendation comprises two key stages: item tokenization and autoregressive generation. Existing item tokenization methods mainly include the following approaches. Clustering‑based approaches, such as SEATER~\cite{si2024generative} and EAGER~\cite{wang2024eager}, cluster item embeddings to construct identifiers. 
Vector-quantization based approaches encompass residual quantization methods, such as TIGER~\cite{TIGER}, LETTER~\cite{LETTER}, and LC-Rec~\cite{LC_Rec}, which employ RQ-VAE~\cite{zeghidour2021soundstream}, and OneRec~\cite{deng2025onerec,zhou2025onerec}, which employs RQ-KMeans; as well as product quantization~\cite{ge2013optimized} methods, such as RPG~\cite{RPG}.
There are also studies~\cite{tang2025think,dai2025onepiece} that enhance generative recommendation through latent reasoning~\cite{hao2024training}.
However, most existing methods rely on an autoregressive paradigm combined with hierarchical quantization, which imposes unidirectional constraints and error accumulation. To address these limitations, we propose parallel semantic IDs and adopt discrete diffusion as the generative~model.

\subsection{Discrete Diffusion Model}
The discrete diffusion model~\cite{li2025survey,gu2022vector,ouyour,nie2025large,zhu2025llada,niescaling}, as an emerging language modeling architecture, has received widespread attention in recent years.
Built upon a bidirectional Transformer backbone, it is trained through a forward token-masking noise process and a reverse denoising reconstruction, which together enable stronger bidirectional context modeling.
During inference, tokens are generated in parallel, while low-confidence tokens are re-masked and re-predicted, yielding an adaptive and flexible generation order.
LLaDA~\cite{nie2025large} represents the first diffusion-based language model to achieve performance comparable to that of autoregressive models. Its extension, LLaDA-V~\cite{you2025llada}, adapts the framework for visual understanding, and MMaDA~\cite{yang2025mmada} further generalizes it to multimodal understanding and generation.
The follow-up version, LLaDA 1.5~\cite{zhu2025llada}, integrates DPO-based post-training for additional performance gains.
In this work, we build upon the discrete diffusion model and leverage its advantages in bidirectional understanding and adaptive generation to advance generative recommendation.

\begin{figure*}[t]
\centering
\includegraphics[width=1.0\textwidth]{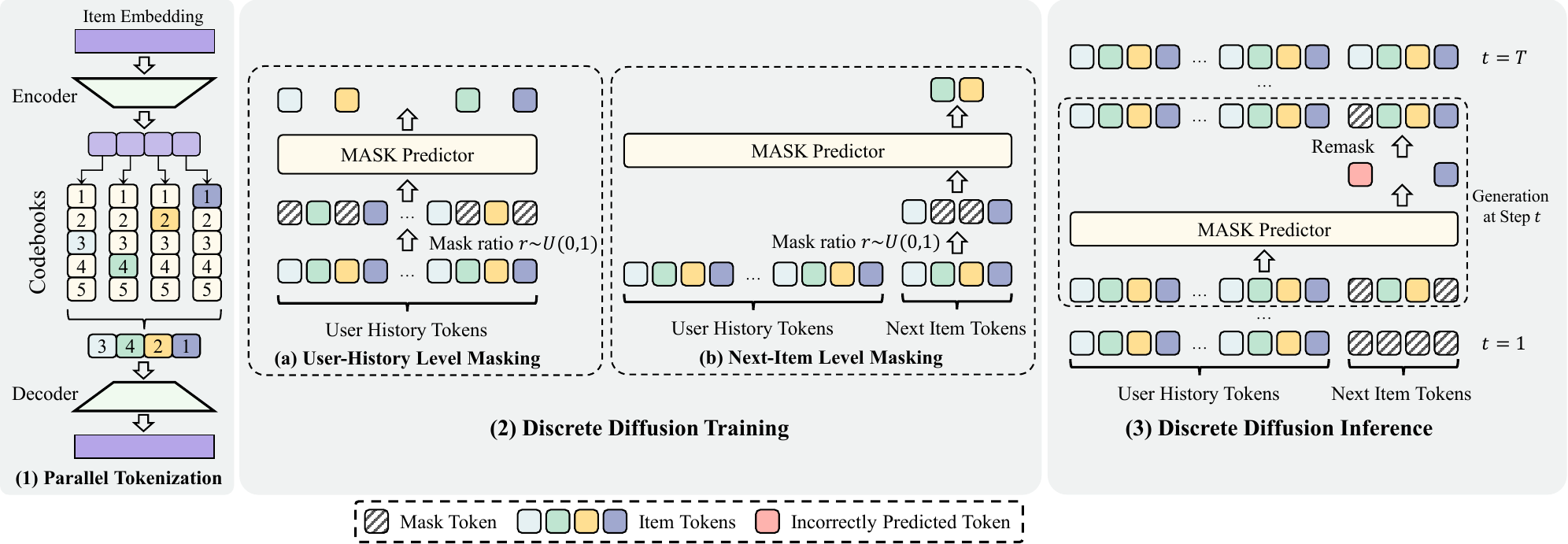}
\caption{
Overall framework of \ourname, which consists of three main modules: 
(1)~Parallel Tokenization, where 
Multi-Head VQ-VAE
are used to produce parallel semantic IDs for each item;
(2)~Discrete Diffusion Training, which applies two masking strategies. The user-history level masking models inter-item sequential dependencies, and the next-item level masking captures intra-item semantic relationships; 
(3)~Discrete Diffusion Inference, where beam search is adapted to discrete diffusion decoding to generate the final top-$k$ recommended items.
}
\label{fig:method}
\end{figure*}

\section{Preliminaries and Background}
This section provides the problem formulation and describes the probability distribution underlying the discrete diffusion model.

\subsection{Problem Formulation}
Let $\mathcal{U}$ and $\mathcal{I}$ denote the sets of users and items, respectively. Each user $u \in \mathcal{U}$ has a chronologically ordered interaction history 
$\mathcal{H} = \{i_1, i_2, \ldots, i_{n-1}\}$, where $i_j$ is the $j$-th interacted item. 
The objective is to predict the next item $i_n$ that user $u$ is likely to interact with.

In general, under the generative recommendation framework, each item is tokenized into a sequence of discrete tokens, namely semantic IDs (SIDs). The SID of item $i$ is denoted as
$s_{i} = [ c_{i,1}, c_{i,2}, \dots, c_{i,M} ]$,
where $M$ is the length of the SID and $c_{i,j}$ is the $j$-th token.
Furthermore, each item in the user’s interaction history can be represented by its SID, resulting in the following token sequence for the user history:
\begin{equation}
\mathcal{S}_{\mathcal{H}} = [c_{1,1}, \dots, c_{1,M},\; 
c_{2,1}, \dots, c_{2,M},\; 
\dots,\; 
c_{n-1,1}, \dots, c_{n-1,M}].
\end{equation}
The task of predicting the next item can be formulated as the following conditional probability maximization:
\begin{equation}
\label{eq:condition_prob}
    \theta^{*} = \underset{\theta}{\text{arg} \max} \mathrm{P}_{\theta}(s_n | \mathcal{S}_{\mathcal{H}}),
\end{equation}
where $\theta$ represents the parameters of the generative model. 
$\theta^{*}$ denotes the optimal parameter.

In the following, we first revisit the probabilistic formulation of the autoregressive approach, and then present the probabilistic formulation of discrete diffusion, which allows us to highlight the differences between the two methods.

\subsection{Autoregressive Modeling}
Existing generative recommendation methods (e.g., TIGER~\cite{TIGER}) predominantly adopt the autoregressive paradigm, generating tokens sequentially from left to right. At each step, the generation is conditioned on both the user’s historical interactions and the tokens generated in preceding steps. Formally, the conditional probability can be expressed as:
\begin{equation}
\label{eq:prob_autoregressive}
\mathrm{P}_{\theta}(s_n \mid \mathcal{S}_{\mathcal{H}}) = 
\prod_{m=1}^{M} \mathrm{P}_{\theta}(c_{n,m} \mid c_{n, <m},\mathcal{S}_{\mathcal{H}}),
\end{equation}
where $c_{n,<m}$ denotes all tokens preceding $c_{n,m}$; for the first token, it corresponds to a special \texttt{[BOS]} token.

\subsection{Discrete Diffusion Modeling}
\label{sec:diffusion_prob}
In contrast to the autoregressive paradigm, generation in discrete diffusion corresponds to the reverse process, in which 
a Mask Predictor
gradually predicts the $M$ tokens of the next item starting from a fully masked sequence containing $M$ \texttt{[MASK]} tokens. At each step, the model simultaneously predicts all positions, retains the token prediction for the position with the highest confidence, and re-masks the remaining positions; the updated sequence is then fed back into the model for further prediction.
Formally, this process can be expressed as:
\begin{equation}
\label{eq:prob_discrete_diffusion}
\mathrm{P}_{\theta}(s_n \mid \mathcal{S}_{\mathcal{H}}) =
\prod_{t=1}^{T} \prod_{m=1}^{M}
\begin{cases}
\mathrm{P}_{\theta}\!\left(c_{n,m} \mid s_{n}^t, \mathcal{S}_{\mathcal{H}}\right), & \text{if } c_{n,m}^t = \texttt{[MASK]}, \\[0.5em]
1, & \text{otherwise}.
\end{cases}
\end{equation}
where $T$ denotes the number of generation steps.
Here, ${s}^t_{n}$ represents the input at step $t$, with the initial state ${s}^1_{n}$ consisting entirely of $M$ \texttt{[MASK]} tokens. 
$c_{n,m}^t$ denotes the $m$-th token of $s_{n}^t$.
At each step, ${s}^t_{n}$ is concatenated with $\mathcal{S}_{\mathcal{H}}$ and passed to the mask predictor, implemented as a Transformer encoder with bidirectional attention.

Unlike Eq.~\eqref{eq:prob_autoregressive}, which requires exactly $M$ steps to generate all tokens sequentially, the generation process in Eq.~\eqref{eq:prob_discrete_diffusion} proceeds for $T$ steps ($T \leq M$). At each step, 
the mask predictor predicts all masked tokens parallel.
the top-$\frac{M}{T}$ tokens with the highest predicted probabilities are retained while the rest are re-masked. 
Overall, the advantages of our approach lie in parallel generation, adaptive generation order, and explicit control over the number of generation~steps.

\section{Our Approach: \ourname}
This section introduces our proposed framework, \ourname, as illustrated in Figure~\ref{fig:method}. Our approach consists of three main modules:
(1)~Parallel Tokenization (Section~\ref{sec:item_tokenize}): we employ 
Multi-Head VQ-VAE
to generate parallel semantic IDs;
(2) Discrete Diffusion Training (Section~\ref{sec:diffusion_train}): we design a User‑History level masking and a Next‑Item level masking to guide the training process;
(3) Discrete Diffusion Inference (Section~\ref{sec:diffusion_infer}): we enhance beam search so that discrete diffusion can generate the top‑$k$ recommended items.

\subsection{\mbox{Parallel Tokenization via Multi-Head VQ-VAE}}
\label{sec:item_tokenize}
To align with the autoregressive generation paradigm, existing generative recommendation models typically employ hierarchical quantization methods such as RQ-VAE~\cite{TIGER,LETTER,LC_Rec} or RQ-KMeans~\cite{deng2025onerec,zhou2025onerec}.
In these frameworks, earlier tokens (e.g., the first token) play a more dominant role, as subsequent tokens are conditionally dependent on them.
In contrast, the bidirectional Transformer architecture adopted in discrete diffusion enables full mutual interactions among all tokens, rendering each token equally important in the representation and generation process.
To better align with this property, we propose a 
\textbf{Multi-Head VQ-VAE}
architecture that eliminates hierarchical dependencies and models all tokens on an equal footing.

Specifically, we first encode the textual information of item \(i\) (e.g., title, description) using a pre-trained embedding model, such as BERT~\cite{devlin2019bert} or Sentence-T5~\cite{ni2021sentence}, to obtain its semantic representation \(\mathbf{v}_i \in \mathbb{R}^D\).  
We then project \(\mathbf{v}_i\) into a latent space via an encoder for subsequent quantization:
\begin{equation}
\label{eq:vq_encoder}
\mathbf{z}_i = \mathrm{Encoder}(\mathbf{v}_i),
\end{equation}
where \(\mathbf{z}_i \in \mathbb{R}^d\) and \(\mathrm{Encoder}(\cdot)\) is implemented as a multilayer perceptron (MLP).
Next, we partition \(\mathbf{z}_i\) into \(M\) sub-vectors:
\begin{equation}
\mathbf{z}_i = [\mathbf{z}_{i,1}; \mathbf{z}_{i,2}; \ldots; \mathbf{z}_{i,M}],
\end{equation}
where \(\mathbf{z}_{i,m} \in \mathbb{R}^{d/M}\) and \(m \in \{1,2,\ldots,M\}\). 
Each sub-vector corresponds to a head and is quantized independently.

We maintain \(M\) codebooks, where the \(m\)-th codebook is defined as
$\mathcal{C}_m = \{\mathbf{e}_{m,k}\}_{k=1}^{K}$,
with \(K\) representing the size of the codebook and \(\mathbf{e}_{m,k} \in \mathbb{R}^{d/M}\) being a learnable code embedding.  
The quantization for the \(m\)-th sub-vector is formulated as:
\begin{equation}
\label{eq:vq_quantization}
c_{i,m} = \arg\min_k \|\mathbf{z}_{i,m} - \mathbf{e}_{m,k}\|_2^2, \quad \mathbf{e}_{m,k} \in \mathcal{C}_m,
\end{equation}
where \(c_{i,m}\) denotes the selected code from the \(m\)-th codebook.

After quantizing all \(M\) sub-vectors, we obtain the semantic ID of item \(i\) as $s_{i} = [ c_{i,1}, c_{i,2}, \dots, c_{i,M} ]$
, along with the corresponding code embeddings \(\{\mathbf{e}_{c_{i,1}}, \mathbf{e}_{c_{i,2}}, \ldots, \mathbf{e}_{c_{i,M}}\}\).  
By concatenating these embeddings, we construct the quantized representation:
\begin{equation}
\hat{\mathbf{z}}_i = [\mathbf{e}_{c_1}; \mathbf{e}_{c_2}; \ldots; \mathbf{e}_{c_M}],
\end{equation}
which is then passed through a decoder to reconstruct the original semantic vector \(\mathbf{v}_i\):
\begin{equation}
\label{eq:vq_decoder}
\hat{\mathbf{v}}_i = \mathrm{Decoder}(\hat{\mathbf{z}}_i),
\end{equation}
where \(\mathrm{Decoder}(\cdot)\) is also implemented as an MLP.

The overall VQ-VAE loss consists of the reconstruction loss $\mathcal{L}_{\mathrm{Recon}}$ and the vector quantization loss $\mathcal{L}_{\mathrm{VQ}}$:
\begin{equation}
\label{eq:vq_loss}
\begin{aligned}
&\mathcal{L}_{\mathrm{Recon}} = \|\mathbf{v}_i - \hat{\mathbf{v}}_i\|_2^2, \\
&\mathcal{L}_{\mathrm{VQ}} = \sum_{m=1}^{M} \Big( \|\mathrm{sg}[\mathbf{z}_{i,m}] - \mathbf{e}_{c_{i,m}}\|_2^2 + \alpha \|\mathbf{z}_{i,m} - \mathrm{sg}[\mathbf{e}_{c_{i,m}}]\|_2^2 \Big), \\
&\mathcal{L}_{\mathrm{VQ-VAE}} = \mathcal{L}_{\mathrm{Recon}} + \mathcal{L}_{\mathrm{VQ}},
\end{aligned}
\end{equation}
where \(\mathrm{sg}[\cdot]\) denotes the stop-gradient operation and \(\alpha\) is a hyper-parameter.
$\mathcal{L}_{\mathrm{Recon}}$ ensures that the reconstructed vector $\hat{\mathbf{v}}_i$ matches the original semantic vector $\mathbf{v}_i$. 
$\mathcal{L}_{\mathrm{VQ}}$ minimizes the distance between each sub-vector and its corresponding code embedding.

\subsection{Discrete Diffusion Training}
\label{sec:diffusion_train}
The original discrete diffusion model is trained for language tasks~\cite{nie2025large}; however, there exists a gap between language and recommendation tasks, as the original model does not inherently understand the semantic IDs used in recommendation. 
To bridge this gap, we design two masking mechanisms: 
(1) a User-History level masking that enables the model to capture inter-item sequential dependencies; and 
(2) a Next-Item level masking that helps the model learn the relationships among tokens within the same item (intra-item), enabling it to generate the next item based on historical context.

\subsubsection{Discrete Diffusion Process}
The discrete diffusion model operates in two complementary stages: the \textbf{forward process} and the \textbf{reverse process}. 
In the forward process, tokens in the input sequence are progressively masked. 
At the extreme masking ratio $r = 1$, all tokens are replaced with \texttt{[MASK]}. 
For intermediate ratios $r \in (0,1)$, the sequence is partially masked, with each token independently having a probability $r$ of being masked and a probability $1 - r$ of remaining visible.
In the reverse process, as $r$ decreases from $1$ to $0$, the model incrementally reconstructs the original sequence, starting from the fully masked state and progressively filling in the masked tokens.

Based on this framework, we design two diffusion mask training strategies: user-history level masking and next-item level masking, which enable our discrete diffusion model to capture both inter-item and intra-item dependencies.

\subsubsection{User-History Level Masking}
\label{sec:his_diffusion}
We first apply the discrete diffusion masking process to the token sequence of the user history, $\mathcal{S}_{\mathcal{H}}$. 
The objective of this procedure is to enable the MASK predictor to effectively capture the global dependencies among all tokens within the user's interaction history.
    
At each diffusion step $r \in (0,1)$, each token in $\mathcal{S}_{\mathcal{H}}$ is independently masked with probability $r$ and remains visible with probability $1-r$. 
The partially masked sequence at step $r$, denoted by $\mathcal{S}_{\mathcal{H}}^r$, is then passed to the MASK predictor, which aims to reconstruct the masked tokens. 
The training loss for this masking strategy is defined as:
\begin{small}
\begin{equation}
\label{eq:loss_his_mask}
\begin{aligned}
\mathcal{L}_{\mathrm{His\text{-}Mask}} = - \mathbb{E}_{r,\mathcal{S}_{\mathcal{H}},\mathcal{S}_{\mathcal{H}}^r} \left[ \frac{1}{r} \sum_{i=1}^{M \times (n-1)} \mathbbm{1}\!\left[\mathcal{S}_{\mathcal{H},i}^r = \texttt{[MASK]} \right] \log \mathrm{P}_{\theta} (\mathcal{S}_{\mathcal{H},i} \mid \mathcal{S}_{\mathcal{H}}^r) \right],
\end{aligned}
\end{equation}
\end{small}
where $\mathbbm{1} \!\left[\mathcal{S}_{\mathcal{H},i}^r = \texttt{[MASK]}\right]$ denotes an indicator function that returns $1$ if the $i$‑th token in the user history $\mathcal{S}_{\mathcal{H}}$ is masked at step $r$, and $0$ otherwise.

\subsubsection{Next-Item Level Masking}
\label{sec:next_item_diffusion}
We next apply the discrete diffusion masking process to the token sequence of the next item, while keeping the user history sequence $\mathcal{S}_{\mathcal{H}}$ fully visible. 
In this setting, the $M$ tokens of the next item are progressively masked across diffusion steps. 
This strategy encourages the MASK predictor to capture the semantic relationships among different tokens within the same item, while generating the next item in a manner conditioned on the historical context.

At each step $r \in (0,1)$, each token in the next item sequence ${s}_{n}$ is independently masked with probability $r$ and remains visible with probability $1-r$. 
The partially masked sequence at step $r$, denoted as ${s}_{n}^{r}$, is then concatenated with the historical tokens $\mathcal{S}_{\mathcal{H}}$ and fed into the MASK predictor to reconstruct the masked tokens. 
The training objective is defined as:
\begin{small}
\begin{equation}
\label{eq:loss_item_mask}
\mathcal{L}_{\mathrm{Item\text{-}Mask}} = - \mathbb{E}_{r,{s}_{n},{s}_{n}^r} \left[ \frac{1}{r} \sum_{i=1}^{M} \mathbbm{1}\!\left[c_{n,i}^r = \texttt{[MASK]} \right] \log \mathrm{P}_{\theta} \!\left({c}_{n,i} \mid {s}_{n}^r, \mathcal{S}_{\mathcal{H}}\right) \right],
\end{equation}
\end{small}
where $\mathbbm{1}\!\left[c_{n,i}^r = \texttt{[MASK]}\right]$ denotes an indicator function that returns $1$ if the $i$‑th token of the next item $s_n$ is masked at step~$r$, and $0$~otherwise.

The loss function in Eq.~\eqref{eq:loss_item_mask} can be shown to be an upper bound on the negative log-likelihood of the conditional model distribution in Eq.~\eqref{eq:condition_prob}, as proven in previous studies~\cite{shi2024simplified,ouyour}:
\begin{equation}
\label{eq:loss_prob_bound}
-\mathbb{E}\!\left[\log \mathrm{P}_{\theta}\left(s_n \mid \mathcal{S}_{\mathcal{H}}\right)\right]
\;\le\;
\mathcal{L}_{\mathrm{Item\text{-}Mask}}.
\end{equation}
Consequently, minimizing the loss in Eq.~\eqref{eq:loss_item_mask} is equivalent to maximizing the conditional probability defined in Eq.~\eqref{eq:condition_prob}.
This validates the soundness of the loss function.

\subsubsection{Joint Training}
To jointly optimize the learning objectives, we train the MASK predictor by combining the loss in Eq.~\eqref{eq:loss_his_mask} with that in Eq.~\eqref{eq:loss_item_mask}. This joint objective encourages the model to capture the semantic information of diverse tokens within the user history, while simultaneously predicting the tokens of the next item conditioned on the historical context. The overall training loss is formulated as:
\begin{equation}
\label{eq:loss_total}
\mathcal{L}_{\mathrm{Total}} = \mathcal{L}_{\mathrm{Item\text{-}Mask}} + \lambda_{\mathrm{His\text{-}Mask}} \mathcal{L}_{\mathrm{His\text{-}Mask}} + \lambda_{\mathrm{Reg}} \|\theta\|_2^2,
\end{equation}
where $\lambda_{\mathrm{His\text{-}Mask}}$ is a weighting coefficient that balances the contributions of $\mathcal{L}_{\mathrm{Item\text{-}Mask}}$ and $\mathcal{L}_{\mathrm{His\text{-}Mask}}$. $\lambda_{\mathrm{Reg}}$ controls the strength of the $L_{2}$ regularization term.

\subsection{Discrete Diffusion Inference}
\label{sec:diffusion_infer}
After training the discrete diffusion model, our objective is to generate the top‑$k$ recommended items. However, we face two challenges.
(1)~The original discrete diffusion models for language generation produce outputs through sampling~\cite{nie2025large}, and their greedy top-$1$ sampling cannot yield top-$k$ results.
(2)~The beam search algorithm, which is commonly used in generative recommendation~\cite{TIGER,LC_Rec} to generate top‑$k$ results, relies on a fixed left‑to‑right generation order. In contrast, discrete diffusion employs an adaptive and dynamically changing generation order, making conventional beam search unsuitable for this setting.

To overcome these challenges, we adapt beam search to the discrete diffusion framework, enabling it to generate the top‑$k$ recommended items. This section introduces our proposed inference~method.

\subsubsection{Initialization}
As shown in Eq.~\eqref{eq:prob_discrete_diffusion}, the generation process is divided into $T$ discrete steps. 
We denote by $\mathcal{PG}_t$ the set of positions that have already been generated at step $t \in \{1,\ldots,T\}$. 
At the first step, $\mathcal{PG}_1 = \varnothing$.  
Let $s_n^t$ denote the token sequence for the next item to be generated at step~$t$.
Specifically, at $t=1$, we initialize $s_n^1 = \{\texttt{[MASK]}, \ldots, \texttt{[MASK]}\}$, consisting of $M$ \texttt{[MASK]} tokens.

Given $s_n^t$ and the user history token sequence $\mathcal{S}_{\mathcal{H}}$ at step~$t$, the MASK predictor outputs a probability distribution over the vocabulary for each masked position:
\begin{small}
\begin{equation}
\begin{aligned}
\mathrm{P}_{\theta}^{t,m} (w \mid s_n^t, \mathcal{S}_{\mathcal{H}}) \in [0,1], \quad
m \in  \{1, \dots, M\} \setminus \mathcal{PG}_t, \quad
w \in \{1, \dots, |\mathcal{W}|\},
\end{aligned}
\end{equation}
\end{small}
where $m$ indexes masked positions and $w$ indexes candidate tokens in the vocabulary.
Here, $\mathcal{W}$ denotes the vocabulary containing all possible tokens, and $\mathrm{P}_\theta^{t,m}(\cdot)$ represents the probability distribution over the vocabulary for position $m$ at step~$t$.

\subsubsection{Generation Position Selection}
Unlike left-to-right autoregressive generation, discrete diffusion predicts all \texttt{[MASK]} positions in parallel at each step. 
To generate tokens iteratively, we first determine the positions to be generated at step~$t$.  
Given that $M$ tokens need to be generated in total over $T$ steps,  
at each step we select the top-$\frac{M}{T}$ unfilled positions with the highest maximum token probabilities:
\begin{equation}
\label{eq:position_select}
\begin{aligned}
\mathcal{M}_{t} = \underset{m \in \{1, \dots, M\} \setminus \mathcal{PG}_t}{\operatorname{top}\text{-}\frac{M}{T}} 
&\left( \max_{w \in \{1, \dots, |\mathcal{W}|\}} \mathrm{P}_{\theta}^{t,m}(w \mid s_n^t, \mathcal{S}_{\mathcal{H}}) \right),\\
\mathcal{PG}_{t+1} &= \mathcal{PG}_{t} \cup \mathcal{M}_{t}.
\end{aligned}    
\end{equation}
Here, $\mathcal{M}_{t}$ denotes the set of positions with the top-$\frac{M}{T}$ confidence scores at step~$t$. 
Following this selection, $\mathcal{PG}_t$ is updated via a union operation to obtain $\mathcal{PG}_{t+1}$.

\subsubsection{Beam Search for Discrete Diffusion}
We perform beam search sequentially over all selected positions in $\mathcal{M}_t$.  
Let $\mathcal{B}_t$ denote the beam set at step~$t$, and 
$\mathcal{B}_{t,0}$ as its initial state before processing any positions in $\mathcal{M}_t$.  
Index the positions in $\mathcal{M}_t$ as $\{ m_1, m_2, \dots, m_{|\mathcal{M}_t|} \}$.
For each $m_i$, we first expand the current beam set with the top-$B$ candidate tokens at that position, and then prune it back to the top-$B$ beams according to the model scores:
\begin{equation}
\begin{aligned}
&\mathcal{B}_{t,0} \; \leftarrow \; \mathcal{B}_t,\quad
\mathcal{B}_{t,i} \; \leftarrow \; \mathcal{B}_{t,i-1} \cup 
    \underset{w \in  \{1, \dots, |\mathcal{W}|\}}{\operatorname{top}\text{-}B}
    \big( \mathrm{P}_{\theta}^{t,m_i}(w \mid s_n^t, \mathcal{S}_{\mathcal{H}}) \big),\\
&\mathcal{B}_{t,i} \; \leftarrow \; 
    \underset{b \in  \mathcal{B}_{t,i}}{\operatorname{top}\text{-}B}
    \big( \mathrm{P}_{\theta}^{t}(b \mid s_n^t, \mathcal{S}_{\mathcal{H}}) \big),\quad
\mathcal{B}_{t+1} \; \leftarrow \; \mathcal{B}_{t,\,|\mathcal{M}_t|}.
\end{aligned}
\end{equation}
Here, $B$ is the beam size, $\operatorname{top}\text{-}B(\cdot)$ selects the $B$ elements with the highest confidence scores, 
and $\mathrm{P}_\theta^{t}(\cdot)$ denotes the joint probability of a beam $b$ at step~$t$.
After beam expansion, the tokens at all selected positions $\mathcal{M}_t$ in $s_n^t$  
are replaced with the newly generated tokens to form the updated sequence $s_n^{t+1}$.

\subsubsection{Iterative Generation}
In each iteration, $\mathcal{M}_t$ is determined according to Eq.~\eqref{eq:position_select}.  
All unselected positions are \textbf{remasked} so that the MASK predictor can re-evaluate them in the context of the partially generated sequence.  
With this updated sequence, the MASK predictor is applied again, and the process of position selection and beam search is repeated until all positions have been filled.  
The resulting sequences are then ranked by their overall probabilities, and the top-$k$ sequences are returned as recommendations.

This iterative procedure enables the discrete diffusion model to refine predictions at unselected positions dynamically, achieving high-quality top-$k$ outputs while providing greater flexibility than strictly left-to-right autoregressive generation.

\begin{table}[t]
\centering
\caption{
Comparison of different generative recommendation~methods.
}
\label{tab:comparison}
\resizebox{.98\columnwidth}{!}{
\begin{tabular}
{lcccc}
\toprule
Methods 
& \makecell[c]{Attention Mechanism}
& \makecell[c]{Generation Order} 
& \makecell[c]{Controllable \\ Generation Step}\\
\midrule
TIGER~\cite{TIGER} 
&Causal
& Left2Right  &\textcolor{purple}{\XSolidBrush}  \\
LETTER~\cite{LETTER} 
&Causal
&Left2Right  &\textcolor{purple}{\XSolidBrush} \\
LC-Rec~\cite{LC_Rec} 
& Causal
&Left2Right  &\textcolor{purple}{\XSolidBrush} \\
RPG~\cite{RPG} 
&Causal
&Parallel  &\textcolor{purple}{\XSolidBrush} \\
\hdashline
\ourname 
&Bidirectional
& Adaptive & \textcolor{teal}{\CheckmarkBold} \\
\bottomrule
\end{tabular}
}
\vspace{-0.3cm}
\end{table}

\subsection{Discussion}

\subsubsection{Continuous vs. Discrete Diffusion in Recommendation}
Continuous diffusion models have been widely applied in image generation~\cite{yang2023diffusion,rombach2022high} and sequential recommendation~\cite{DreamRec,DiffuRec,wang2023diffusion,li2025dimerec}. 
Operating in continuous spaces, they produce high-quality images or latent representations through a forward noising process followed by reverse denoising. 
In contrast, discrete diffusion models~\cite{li2025survey,nie2025large} are designed to generate sequences of discrete tokens.

In recommendation tasks, continuous-diffusion-based approaches generally produce a latent representation which is later used to retrieve items via similarity search. 
In such methods, retrieval and model optimization are conducted as separate stages. 
By comparison, our discrete diffusion approach directly generates the semantic IDs of items, thereby removing the retrieval stage and unifying generation and retrieval into a single optimization process. 
This integration simplifies the inference pipeline and leads to improved recommendation performance.

\subsubsection{Advantages over Autoregressive Models}
Autoregressive generative recommendation methods~\cite{TIGER,LC_Rec,LETTER,ETEGRec} generate tokens in a fixed left-to-right order, resulting in strong sequential dependencies and susceptibility to error accumulation, where an early mistake can propagate through the entire sequence. Their attention mechanism is unidirectional (causal), restricting each token to attend only to its preceding tokens. Although RPG~\cite{RPG} employs multi-token prediction~\cite{gloeckle2024better} for parallel generation of semantic IDs, the prediction is performed in a single step, preventing iterative refinement through re-masking and re-prediction. Moreover, it requires a complex decoding strategy.

In contrast, discrete diffusion uses a adaptive, confidence-driven generation order that prioritizes tokens with high certainty, thereby reducing the impact of early-step errors and alleviating error accumulation. It also incorporates bidirectional attention, allowing tokens to attend to both preceding and succeeding positions to capture richer contextual semantics. 
Furthermore, discrete diffusion supports controllable generation steps, and can predict multiple tokens at each step.

Overall, bidirectional modeling combined with adaptive generation order aligns the generation process more closely with recommendation objectives, improving accuracy. Table~\ref{tab:comparison} summarizes the key differences between \ourname and representative generative recommendation methods.

\begin{table}[t]
\small
\centering
\caption{
Statistics of the used datasets. ``Avg.\emph{len}'' indicates the average number of interactions within each input sequence.
}
\label{tab:dataStatistics}  
 \resizebox{1.0\columnwidth}{!}{
\begin{tabular}{cccccc}
\toprule
Dataset & \#Users & \#Items  & \#Interaction &Sparsity &Avg.\emph{len} \\
\midrule
Scientific &50,985 &25,848 &412,947 &99.969\% &8.10 \\
Instrument &57,439 &24,587 &511,836 &99.964\% &8.91 \\
Game       &94,762 &25,612 &814,586 &99.966\% &8.60 \\
\bottomrule
\end{tabular}
}
\vspace{-0.3cm}
\end{table}

\begin{table*}[h!]
\small
\centering
\caption{
Recommendation performance of various methods across the three datasets. The best-performing and second-best methods are denoted with boldface and underlining, respectively. 
The improvements over the second-best methods are statistically significant (paired $t$-test, $p$-value$<0.05$).
}
\label{tab:main_result}
\resizebox{1.0\linewidth}{!}{
\begin{tabular}{
ll
ccccccc
ccccccc
}
\toprule
\multicolumn{1}{l}{\multirow{2}{*}{Datasets}} & 
\multicolumn{1}{l}{\multirow{2}{*}{Metric}} & 
\multicolumn{7}{c}{Item ID-based} & 
\multicolumn{7}{c}{Semantic ID-based} \\ 
\cmidrule(l){3-9} \cmidrule(l){10-16}
\multicolumn{1}{c}{}  & \multicolumn{1}{c}{} 
& GRU4Rec & SASRec & BERT4Rec &FMLP-Rec & LRURec & DreamRec & DiffuRec 
& VQ-Rec & TIGER &TIGER-SAS & LETTER & LC-Rec
& RPG & \textbf{\ourname}  \\ 
\midrule
\multirow{7} * {Scientific} 
&Recall@1 &0.0071 &0.0063 &0.0045 &0.0046 &0.0049 &0.0052 &0.0050 &0.0076 &0.0084 &0.0067 &0.0082 &\underline{0.0091} &0.0087 &\textbf{0.0098}  \\
&Recall@5 &0.0184 &0.0240 &0.0157 &0.0181 &0.0169 &0.0184 &0.0190 &0.0248 &\underline{0.0282} &0.0221 &0.0273 &0.0280 &0.0257 &\textbf{0.0310} \\
&Recall@10 &0.0272 &0.0379 &0.0264 &0.0300 &0.0267 &0.0299 &0.0310 &0.0385 &\underline{0.0446} &0.0356 &0.0423 &0.0434 &0.0395 &\textbf{0.0474} \\
&NDCG@5 &0.0128 &0.0152 &0.0100 &0.0113 &0.0110 &0.0118 &0.0119 &0.0162 &0.0183 &0.0144 &0.0179 &\underline{0.0186} &0.0174 &\textbf{0.0203} \\
&NDCG@10 &0.0156 &0.0197 &0.0134 &0.0151 &0.0141 &0.0155 &0.0158 &0.0206 &\underline{0.0236} &0.0187 &0.0227 &0.0235 &0.0218 &\textbf{0.0256} \\
\hline
\multirow{7} * {Instrument}
&Recall@1  &0.0094 &0.0089 &0.0065 &0.0086 &0.0071 &0.0069 &0.0077 &0.0099 &0.0105 &0.0102 &0.0114 &\underline{0.0119} &0.0118 &\textbf{0.0128} \\
&Recall@5 &0.0297 &0.0331 &0.0255 &0.0299 &0.0272 &0.0245 &0.0283 &0.0345 &0.0359 &0.0342 &0.0362 &\underline{0.0379} &0.0362 &\textbf{0.0406} \\
&Recall@10 &0.0453 &0.0525 &0.0412 &0.0496 &0.0431 &0.0423 &0.0465 &0.0532 &0.0566 &0.0521 &0.0562 &\underline{0.0587} &0.0545 &\textbf{0.0623} \\
&NDCG@5 &0.0196 &0.0211 &0.0160 &0.0193 &0.0172 &0.0157 &0.0179 &0.0222 &0.0233 &0.0223 &0.0239 &\underline{0.0251} &0.0241 &\textbf{0.0268} \\
&NDCG@10 &0.0246 &0.0273 &0.0211 &0.0257 &0.0223 &0.0214 &0.0237 &0.0282 &0.0300 &0.0280 &0.0303 &\underline{0.0318} &0.0300 &\textbf{0.0337} \\
\hline
\multirow{7} * {Game}
&Recall@1  &0.0149 &0.0128 &0.0082 &0.0099 &0.0134 &0.0125 &0.0111 &0.0150 &0.0166 &0.0170 &0.0169 &0.0165 &\textbf{0.0209} &\underline{0.0203} \\
&Recall@5 &0.0461 &0.0516 &0.0315 &0.0395 &0.0480 &0.0381 &0.0425 &0.0497 &0.0529 &0.0548 &0.0552 &0.0567 &\underline{0.0579} &\textbf{0.0623}  \\
&Recall@10 &0.0712 &0.0823 &0.0530 &0.0649 &0.0753 &0.0611 &0.0709 &0.0769 &0.0823 &0.0847 &0.0863 &\underline{0.0891} &0.0853 &\textbf{0.0942} \\
&NDCG@5 &0.0307 &0.0323 &0.0199 &0.0246 &0.0308 &0.0253 &0.0268 &0.0325 &0.0348 &0.0360 &0.0362 &0.0366 &\underline{0.0397} &\textbf{0.0415}  \\
&NDCG@10 &0.0387 &0.0421 &0.0267 &0.0328 &0.0396 &0.0326 &0.0359 &0.0412 &0.0442 &0.0457 &0.0462 &0.0471 &\underline{0.0485} &\textbf{0.0517}  \\
\bottomrule
\end{tabular} 
}
\vspace{-0.3cm}
\end{table*}

\section{Experiments}
We performed comprehensive experiments to assess the effectiveness of \ourname.
The code is available\footnote{\url{https://github.com/TengShi-RUC/LLaDA-Rec}}.

\subsection{Experimental Setup}

\subsubsection{Dataset}
We evaluate our method on three categories from the widely used Amazon 2023 Review dataset~\cite{hou2024bridging}: ``Industrial Scientific'' (Scientific), ``Musical Instruments'' (Instrument), and ``Video Games'' (Game).
In line with~\cite{TIGER,VQRec}, each user’s historical reviews are treated as interaction records and ordered chronologically, with the earliest review placed first.  
For evaluation, we adopt the widely used leave-one-out protocol~\cite{SASREC,TIGER}: in each user sequence, the last item is held out for testing and the second-to-last item is reserved for validation.  
The detailed statistics of the three datasets are presented in Table~\ref{tab:dataStatistics}.

\subsubsection{Baselines}
We first compared our approach with traditional item ID-based methods.
\textbf{Item ID-based}:
(1)~\textbf{GRU4Rec}~\cite{GRU4REC} leverages gated recurrent units (GRUs) to model user interaction histories;  
(2)~\textbf{SASRec}~\cite{SASREC} employs a unidirectional Transformer to capture sequential dependencies;  
(3)~\textbf{BERT4Rec}~\cite{BERT4REC} utilizes a bidirectional Transformer trained with a cloze-style objective;  
(4)~\textbf{FMLP-Rec}~\cite{FMLPREC} adopts multi-layer perceptrons (MLPs) with learnable filters for sequential modeling;  
(5)~\textbf{LRURec}~\cite{LRURec} integrates linear recurrent units (LRUs) to efficiently process long-range user interactions.
(6)~\textbf{DreamRec}~\cite{DreamRec} uses SASRec outputs in a diffusion denoising module, removing negative sampling and training only on positive samples;
(7)~\textbf{DiffuRec}~\cite{DiffuRec} combines generative diffusion with sequential recommendation via a Transformer approximator to reconstruct target item embeddings.

We also compared our approach with generative recommendation methods based on semantic IDs.
\textbf{Semantic ID-based}:
(8)~\textbf{VQ-Rec}~\cite{VQRec} applies product quantization to tokenize items into semantic IDs, which are then pooled to obtain item representations;
(9)~\textbf{TIGER}~\cite{TIGER} utilizes RQ‑VAE to generate codebook identifiers, embedding semantic information into discrete code sequences;
(10)~\textbf{TIGER-SAS}~\cite{TIGER} derives semantic IDs from SASRec‑trained item embeddings rather than text embeddings;
(11)~\textbf{LETTER}~\cite{LETTER} develops a learnable tokenizer that incorporates hierarchical semantics, collaborative signals, and code assignment diversity;
(12)~\textbf{LC-Rec}~\cite{LC_Rec} exploits identifiers with auxiliary alignment tasks to associate the generated codes with natural language;
(13)~\textbf{RPG}~\cite{RPG} is a lightweight semantic ID‑based model that generates long, unordered semantic IDs in parallel via multi‑token~prediction~\cite{gloeckle2024better}.

\subsubsection{Evaluation Metrics}
Following prior studies~\cite{TIGER,LETTER,LC_Rec}, we evaluate performance using two commonly adopted ranking metrics: top-$k$ \textit{Recall} and top-$k$ \textit{Normalized Discounted Cumulative Gain} (NDCG). 
We report results for $k \in \{1, 5, 10\}$. 
Since NDCG@1 is identical to Recall@1, it is omitted from the evaluation.

\subsubsection{Implementation Details}
For the parallel tokenization module, we adopt Sentence-T5~\cite{ni2021sentence} to encode the title and other textual information of each item into an embedding. 
We employ $M=4$ codebooks, each containing $K=256$ code vectors of dimension $d=32$. 
The weight $\alpha$ in Eq.~\eqref{eq:vq_loss} is set to 0.25. 
The 
Multi-Head VQ-VAE
is trained for 10{,}000 epochs using the AdamW optimizer~\cite{loshchilov2017decoupled} with a learning rate of $1\times 10^{-3}$ and a batch size of 2{,}048.

For the discrete diffusion model, the MASK predictor is a bidirectional Transformer encoder with a token embedding dimension of 256 and 8 attention heads per layer. 
On the Scientific and Instrument datasets, we use a 4-layer encoder, whereas on the Game dataset we use a 6-layer encoder. 
The model parameters are randomly initialized and trained using our designed loss function.
The weight $\lambda_{\mathrm{His\text{-}Mask}}$ in Eq.~\eqref{eq:loss_total} is tuned over $\{1,2,3,4,5\}$. 
We train the Transformer encoder for 150 epochs with early stopping, using the AdamW optimizer. 
The learning rate is tuned over $\{0.005, 0.003, 0.001\}$ and the weight decay over $\{0.05, 0.005, 0.001\}$. 
The batch size is set to 1{,}024.

\subsection{Overall Performance}
Table~\ref{tab:main_result} reports the results across the three datasets, and we can observe that:
\begin{itemize}[leftmargin=*]
    \item Firstly, compared with both traditional item ID-based approaches and generative semantic ID-based approaches, \ourname achieves state-of-the-art (SOTA) results. This confirms the benefits of discrete diffusion training and inference, together with the design of 
    Multi-Head VQ-VAE.
    \item Secondly, we can see that generative recommendation methods based on semantic IDs generally outperform traditional methods based on item IDs, which verifies the advantages of using semantic IDs to capture the correlations between different items as well as the benefits of generative approaches.
    \item 
    Finally, we can also observe that both the parallel semantic ID-based methods, RPG and \ourname, achieve promising results. This confirms the advantages of parallel semantic IDs, and the superior performance of \ourname further demonstrates the benefits of discrete diffusion.
\end{itemize}

\begin{table}[!t]
\small
\centering
\caption{
Ablation results on the three datasets. ``w/o'' indicates that the corresponding module is removed. R@5 and N@5 denote Recall@5 and NDCG@5, respectively.
}
\label{tab:ablation_result}
\renewcommand{\arraystretch}{1.2}
\resizebox{1.0\columnwidth}{!}{
\begin{tabular}
{l
cccccc}
\toprule
\multicolumn{1}{l}{\multirow{2}{*}{Model}} & 
\multicolumn{2}{c}{Scientific} & 
\multicolumn{2}{c}{Instrument} & 
\multicolumn{2}{c}{Game} \\ 
\cmidrule(l){2-3} \cmidrule(l){4-5} \cmidrule(l){6-7}
&R@5 &N@5 &R@5 &N@5 &R@5 &N@5  \\ 
\midrule
\textbf{\ourname} &\textbf{0.0310} &\textbf{0.0203} &\textbf{0.0406} &\textbf{0.0268} &\textbf{0.0623} &\textbf{0.0415} \\
\midrule
\multicolumn{7}{c}{Tokenizer} \\
\hdashline
RQ-VAE &0.0293 &0.0191 &0.0367 &0.0244 &0.0604 &0.0399\\
RQ-Kmeans &0.0250 &0.0165 &0.0344 &0.0224 &0.0552 &0.0370\\
OPQ &0.0237 &0.0155 &0.0340 &0.0229 &0.0552 &0.0362\\
\midrule
\multicolumn{7}{c}{Training} \\
\hdashline
w/o $\mathcal{L}_{\mathrm{His\text{-}Mask}}$ &0.0255 &0.0169 &0.0321 &0.0209 &0.0544 &0.0356 \\
w/o $\mathcal{L}_{\mathrm{Item\text{-}Mask}}$ &0.0264 &0.0172 &0.0355 &0.0231 &0.0571 &0.0376 \\
\midrule
\multicolumn{7}{c}{Inference} \\
\hdashline
w/o Beam Search
&0.0077 &0.0077 &0.0091 &0.0091 &0.0162 &0.0162\\
\bottomrule
\end{tabular}
} 
\vspace{-5px}
\end{table}

\subsection{Ablation Study}
We conducted ablation studies on three datasets to verify the effect of each module in \ourname, as shown in Table~\ref{tab:ablation_result}.

\subsubsection{Tokenizer}
As discussed in Section~\ref{sec:item_tokenize}, to adapt the bidirectional Transformer-based discrete diffusion framework, we design 
Multi-Head VQ-VAE.
We compare these with several commonly used semantic ID generation methods, including RQ‑VAE~\cite{TIGER,LC_Rec}, RQ‑Kmeans~\cite{deng2025onerec}, and OPQ~\cite{VQRec,RPG}.
Experimental results show that semantic IDs derived from residual quantization (RQ) consistently underperform, suggesting that RQ is not well aligned with bidirectional Transformers. This misalignment arises because, in RQ, earlier tokens tend to exert greater influence, whereas in \ourname, token importance is uniformly distributed, making it more suitable for the parallel semantic ID~structure.

Furthermore, even when replacing our tokenizer with RQ‑VAE, our method still surpasses the baseline in most cases, thereby validating the robustness and architectural advantage of our generative framework.
Finally, clustering-based approaches such as RQ‑Kmeans and OPQ exhibit inferior performance compared to RQ‑VAE and VQ‑VAE, confirming that VAE‑based quantization methods possess stronger representational capacity.

\subsubsection{Training}
In Section~\ref{sec:diffusion_train}, we introduce two mask mechanisms for training and analyze the impact of the two corresponding loss functions on the results. 
From the results, we observe that removing either loss function leads to performance degradation. 
The loss $\mathcal{L}_{\mathrm{His\text{-}Mask}}$ (Eq.~\eqref{eq:loss_his_mask}) facilitates better modeling of relationships between different items. 
The loss $\mathcal{L}_{\mathrm{Item\text{-}Mask}}$ (Eq.~\eqref{eq:loss_item_mask}) aids in capturing the semantic relationships among different tokens within the same item, 
and enables the model to predict the next item conditioned on the given history.

\subsubsection{Inference}
As discussed in Section~\ref{sec:diffusion_infer}, the original diffusion language model~\cite{nie2025large}, which relies on probabilistic sampling, returns only the top-1 result. 
We adapt beam search so that it can generate the top-$k$ recommended items. 
To assess its impact, we remove beam search and instead apply the original diffusion language model with a greedy search strategy to return the top-1 result. 
The performance drops substantially, confirming the importance of beam search in generative recommendation tasks. 
It is worth noting that greedy search produces only a single top-1 result, which causes the values of Recall and NDCG to become~identical.

\subsection{Experimental Analysis}
We further performed experimental analysis to assess the effectiveness of \ourname.

\begin{figure}[t]
     \centering
     \subfigure[Different Attention Mechanism]{
        \label{fig:attention_mask}
        \includegraphics[width=0.95\columnwidth]{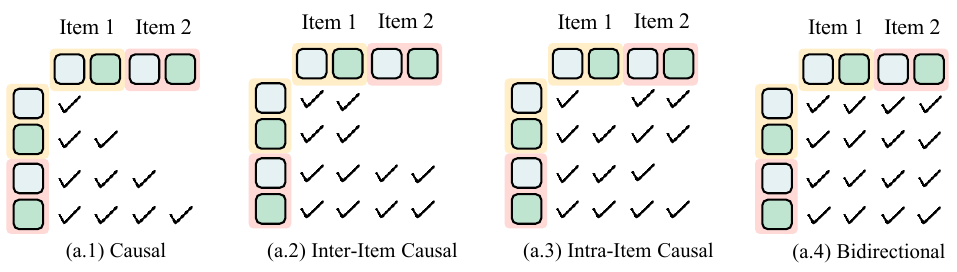}
     }
     \subfigure[Instrument]{
        \label{fig:instrument_attention}
        \includegraphics[width=0.475\columnwidth]{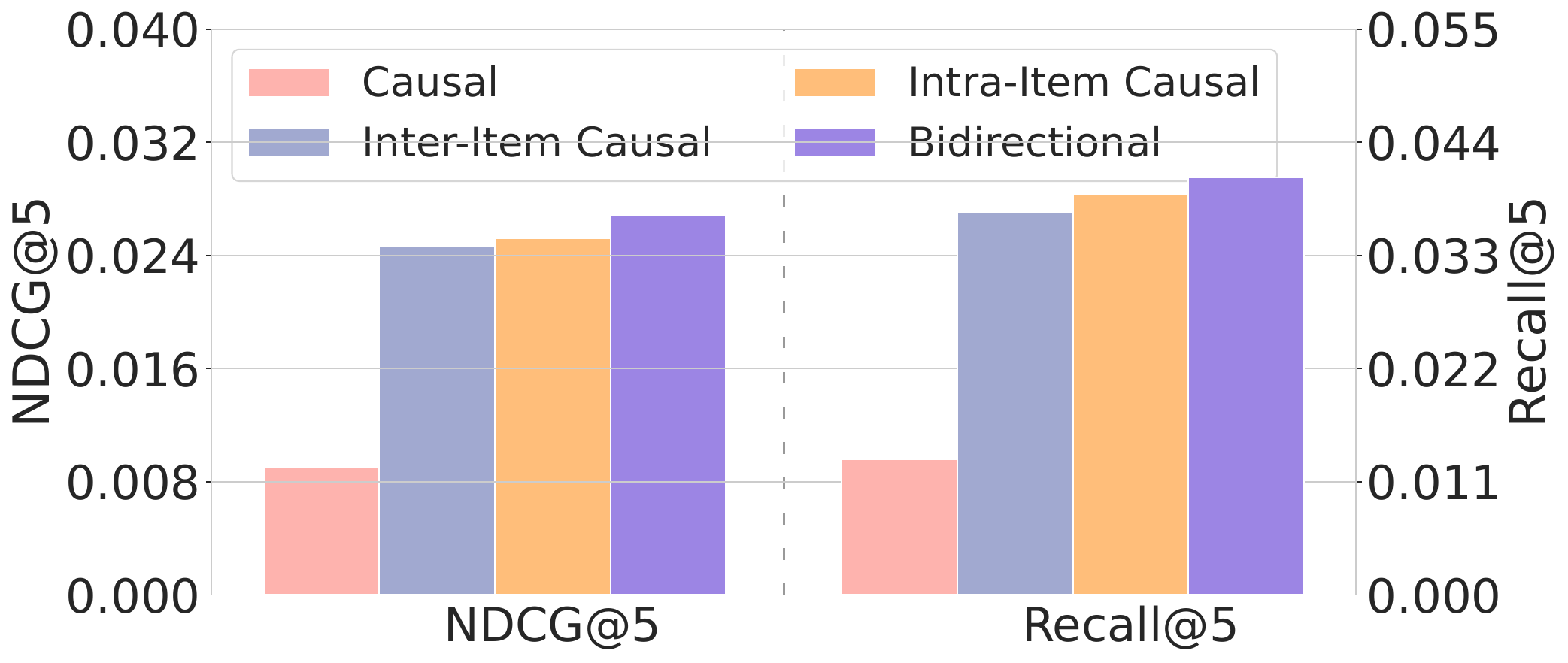}
     }
    \subfigure[Game]{
        \label{fig:game_gen_attention}
        \includegraphics[width=0.475\columnwidth]{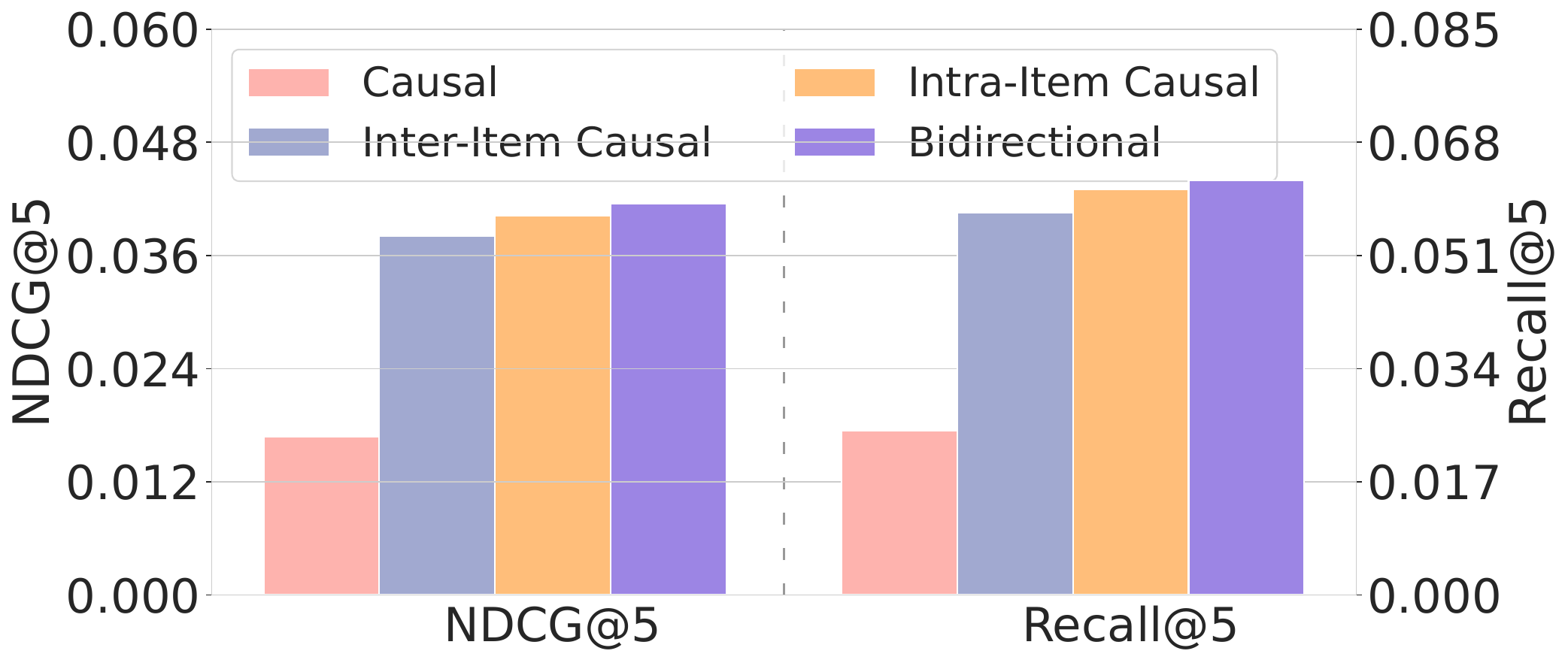}
     }
     \caption{
     Comparison of different attention mechanisms.
    (a):~Attention masks corresponding to each mechanism. 
    (b) and (c):~Performance under different attention mechanisms.
     }
     \label{fig:attention}
\end{figure}

\subsubsection{Impact of the Attention Mechanism}
Our method is built upon a bidirectional Transformer, which offers enhanced capability for modeling contextual dependencies in both directions. 
We compare it against several alternative attention mechanisms:
(1)~causal attention, commonly employed in autoregressive models;
(2)~inter-item causal attention, in which attention across different items is causal, while attention within each item is bidirectional; and
(3)~intra-item causal attention, in which attention within each item is causal, while attention across different items is bidirectional.
The attention masks corresponding to these mechanisms are illustrated in Figure~\ref{fig:attention_mask}.

Figure~\ref{fig:attention} further compares their performance. 
As shown, bidirectional attention yields the best results, attributed to its superior ability to capture contextual dependencies. 
In contrast, causal attention, constrained by its unidirectional structure, exploits contextual information less effectively and thus performs the worst. 
Both inter-item causal and intra-item causal attention achieve competitive performance, 
highlighting that incorporating bidirectional attention—either across items or within items—is crucial for effective contextual modeling.

\begin{figure}[t]
     \centering
     \subfigure[Instrument]{
        \label{fig:instrument_gen_order}
        \includegraphics[width=0.475\columnwidth]{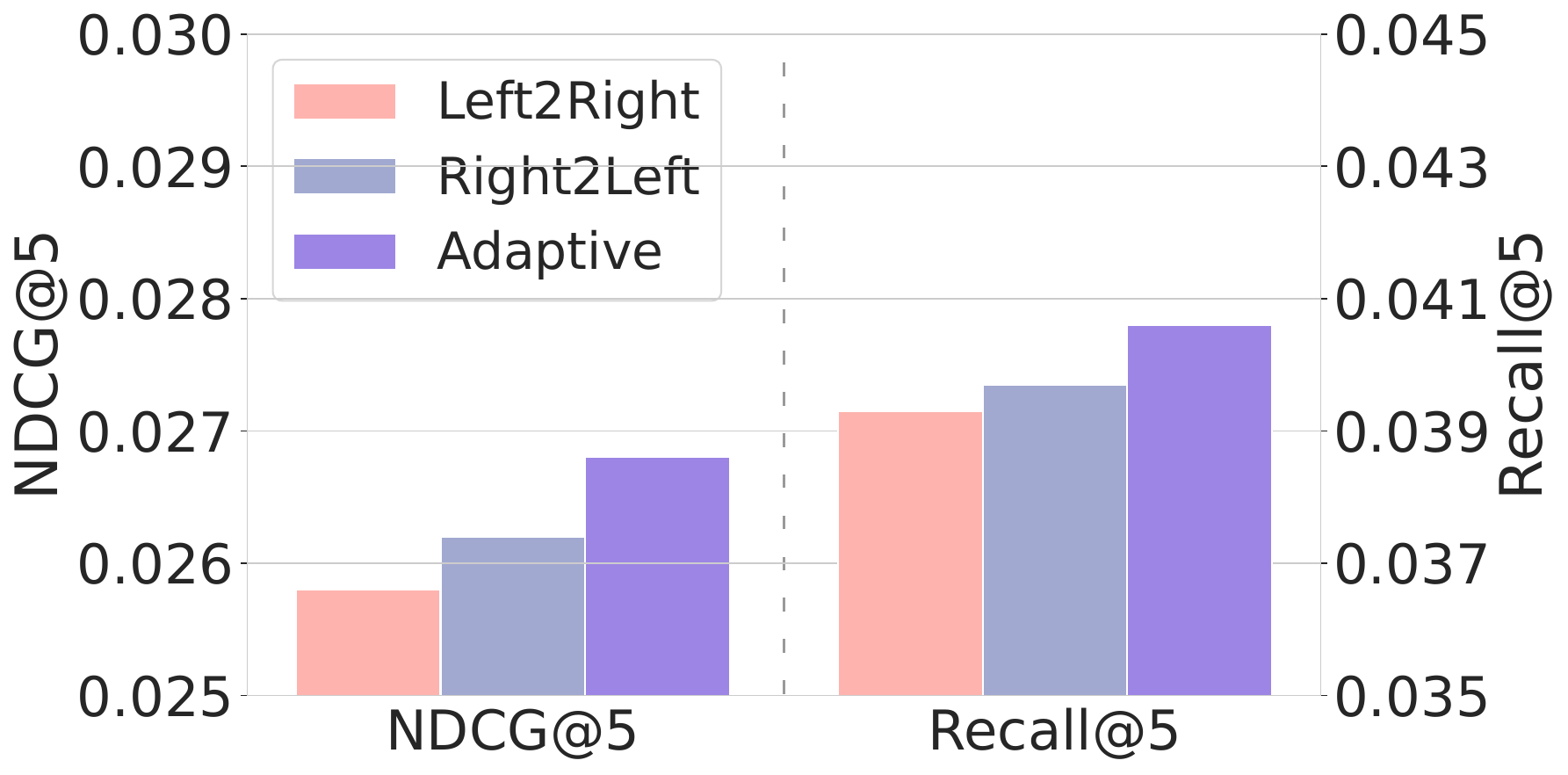}
     }
    \subfigure[Game]{
        \label{fig:game_gen_step}
        \includegraphics[width=0.475\columnwidth]{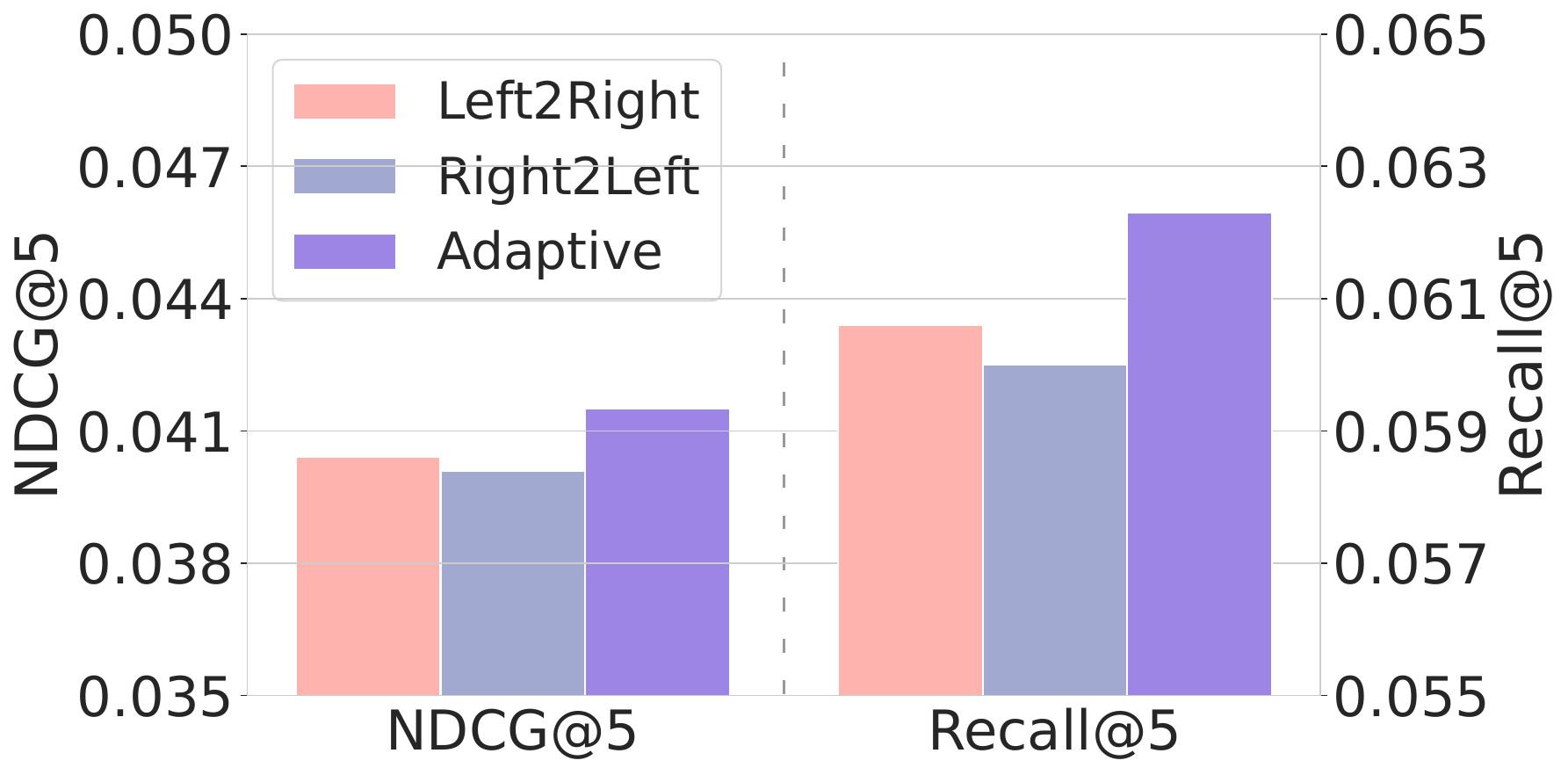}
     }
     \caption{
     Performance under different generation orders.
     }
     \label{fig:gen_order}
\end{figure}

\begin{figure}[t]
     \centering
     \subfigure[Instrument]{
        \label{fig:instrument_gen_step}
        \includegraphics[width=0.475\columnwidth]{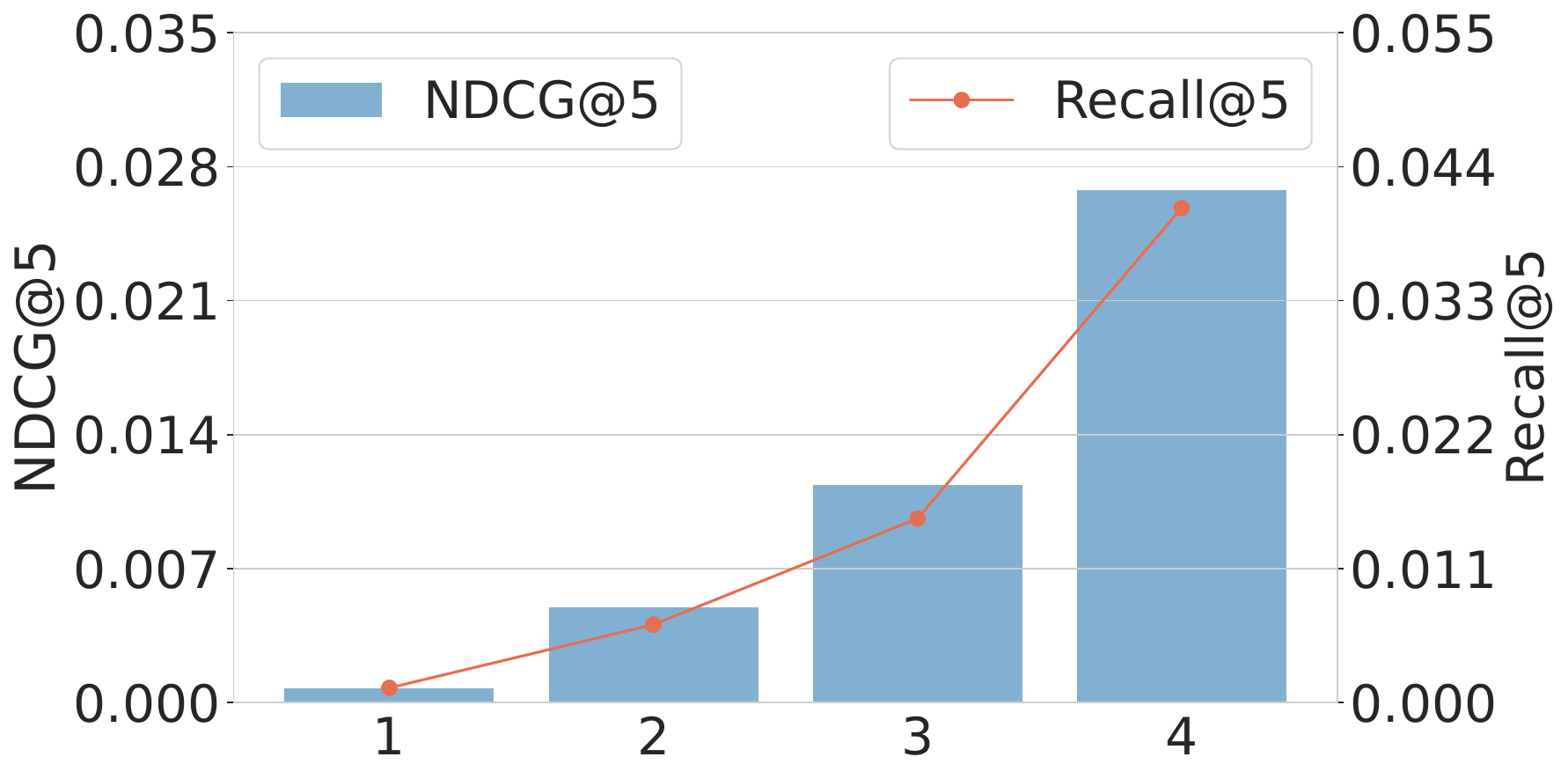}
     }
    \subfigure[Game]{
        \label{fig:game_gen_step}
        \includegraphics[width=0.475\columnwidth]{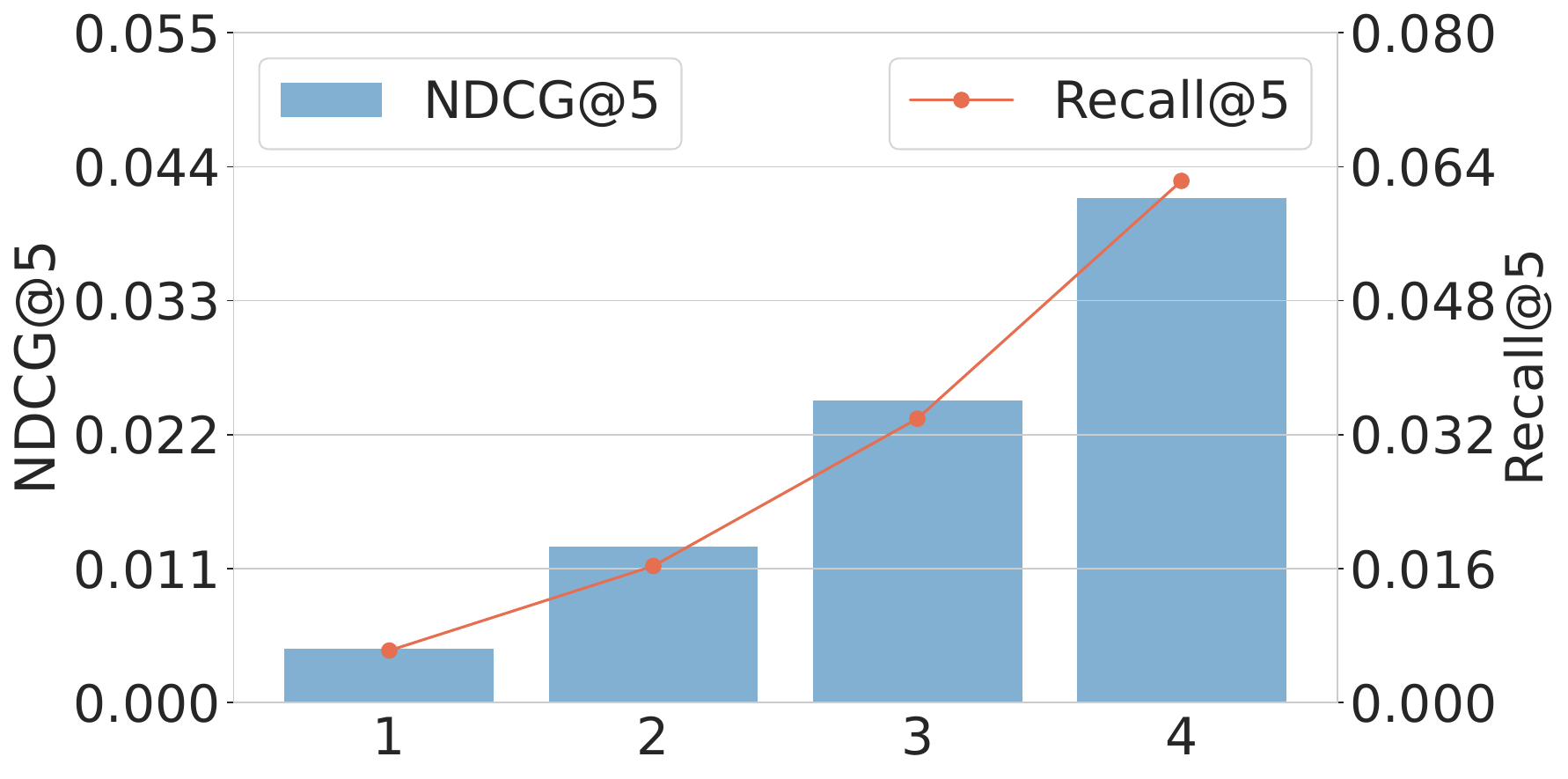}
     }
     \caption{
     Performance under different generation steps.
     }
     \label{fig:gen_step}
\end{figure}

\begin{figure}[t]
     \centering
     \subfigure[Instrument]{
        \label{fig:instrument_pretrain_w}
        \includegraphics[width=0.475\columnwidth]{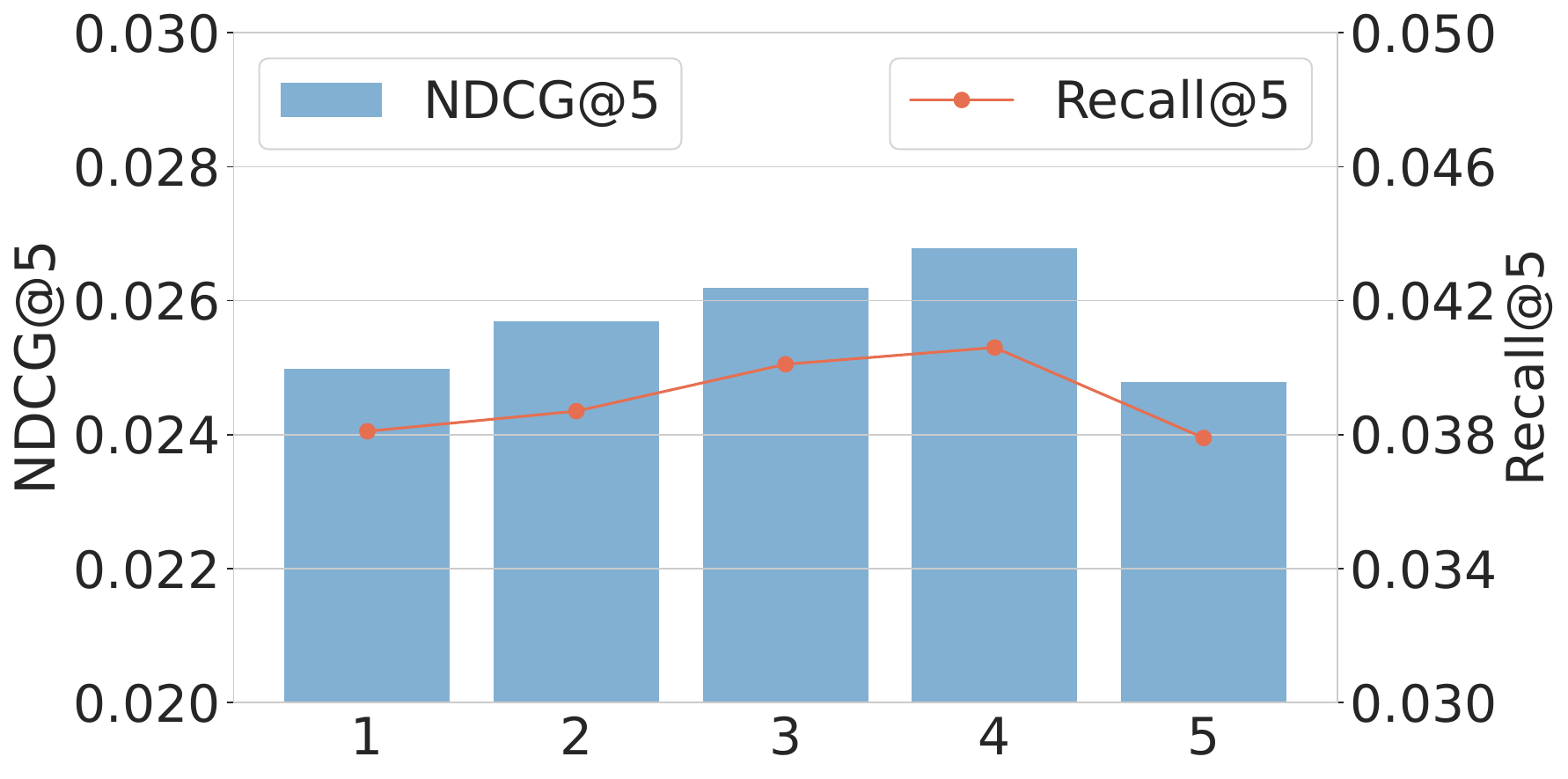}
     }
    \subfigure[Game]{
        \label{fig:game_pretrain_w}
        \includegraphics[width=0.475\columnwidth]{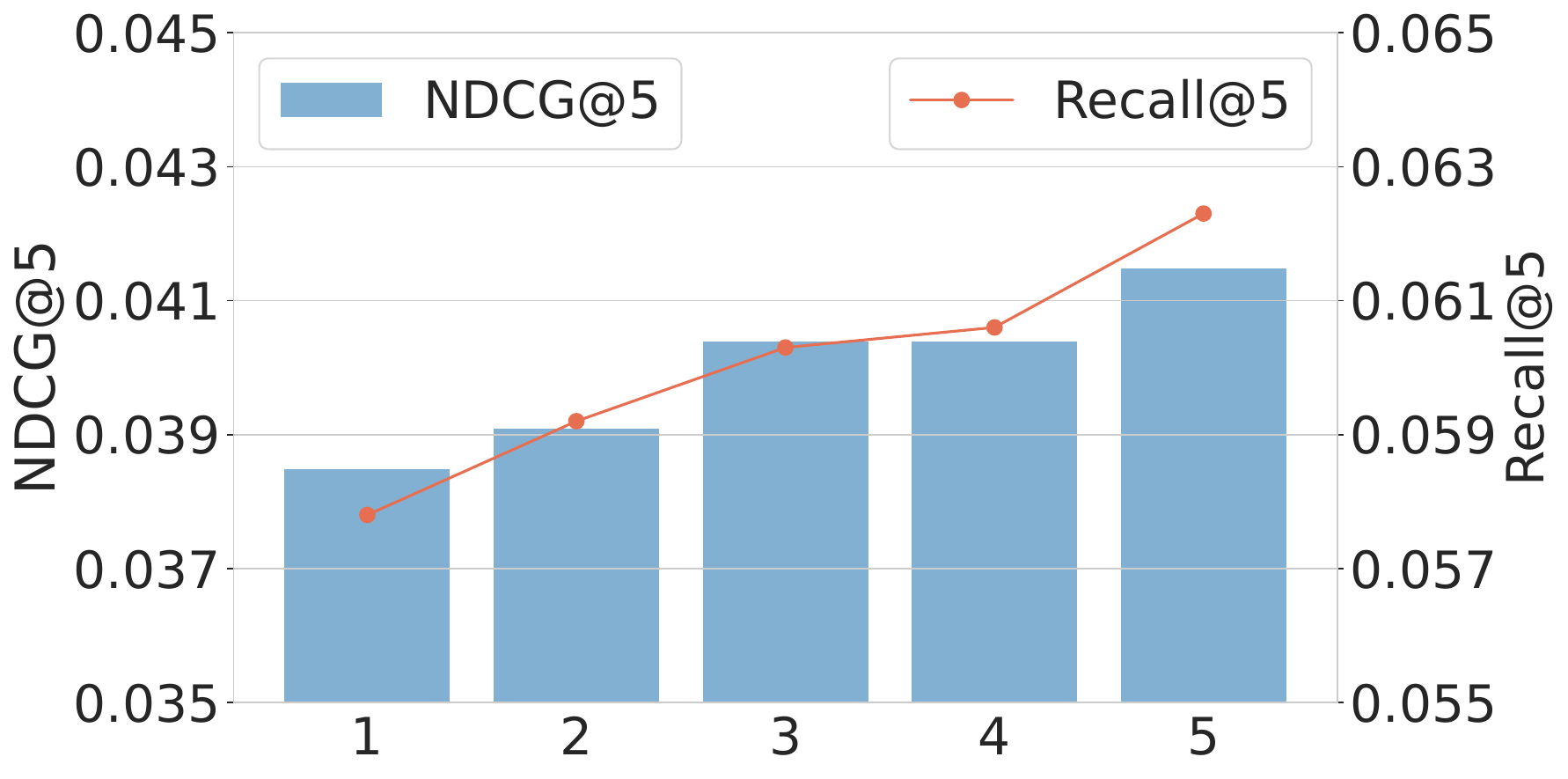}
     }
     \caption{
     Performance of different $\lambda_{\mathrm{His\text{-}Mask}}$ (Eq.~\eqref{eq:loss_total}) values.
     }
     \label{fig:pretrain_w}
\end{figure}

\subsubsection{Impact of Generation Order}
As outlined in Section~\ref{sec:diffusion_infer}, our generation order is determined dynamically by selecting, at each step, the position with the highest model confidence. We compare this adaptive strategy with fixed left-to-right (left2right) and right-to-left (right2left) orders (Figure~\ref{fig:gen_order}). The adaptive approach consistently delivers superior performance, with left2right occasionally producing the poorest results, highlighting the shortcomings of the commonly used left2right scheme. Our method’s ability to progress from easier to 
harder tokens provides a notable advantage.

\subsubsection{Impact of Generation Steps}
As discussed in Sections~\ref{sec:diffusion_prob} and ~\ref{sec:diffusion_infer}, we can control the number of generation steps so that the model produces multiple tokens per step. We analyze the impact of the generation steps on performance, as shown in Figure~\ref{fig:gen_step}. It can be observed that increasing the number of steps leads to better results. Although using fewer steps improves generation efficiency, it also results in a significant performance drop. How to achieve a better trade-off between efficiency and performance with fewer steps remains an open question, and some recent studies on diffusion language models have started to explore this direction~\cite{chen2025dlm,hayakawa2024distillation}.

\subsubsection{Impact of Hyper-parameters}
We investigate the impact of different hyperparameters on the results. 
Specifically, we analyze the effect of the weight $\lambda_{\mathrm{His\text{-}Mask}}$ in Eq.~\eqref{eq:loss_total}, 
as illustrated in Figure~\ref{fig:pretrain_w}. 
The results show that moderately increasing $\lambda_{\mathrm{His\text{-}Mask}}$ can improve performance, 
enabling the model to better capture relationships among different tokens in the history. 
However, setting $\lambda_{\mathrm{His\text{-}Mask}}$ too large may hinder the model’s ability 
to predict the next item conditioned on the given history.

\section{Conclusion}
In this work, we revisit generative recommendation from the perspective of discrete diffusion, addressing the unidirectional constraints and error accumulation inherent in existing autoregressive approaches. 
We propose \ourname, a bidirectional generative recommendation framework that leverages parallel semantic IDs, tailored masking mechanisms, and an adapted beam search strategy to align discrete diffusion with recommendation tasks. 
This design enables the model to capture both global inter-item and intra-item dependencies while mitigating the propagation of early-stage prediction errors. 
Extensive experiments on three real-world datasets demonstrate that \ourname consistently outperforms both traditional ID-based recommenders and state-of-the-art semantic-ID-based generative recommendation models.

\bibliographystyle{ACM-Reference-Format}
\bibliography{ref}

\end{document}